\RequirePackage{lineno}
\documentclass[aps,prl,twocolumn,superscriptaddress,showpacs,preprintnumbers,amsmath,amssymb]{revtex4}

\usepackage{graphicx} % Include figure files
\usepackage{dcolumn}  % Align table columns on decimal point

\graphicspath{{ps}}

\renewcommand{\arraystretch}{1.1}

\usepackage{array}
\newcommand{\PreserveBackslash}[1]{\let\temp=\\#1\let\\=\temp}
\newcolumntype{C}[1]{>{\PreserveBackslash\centering}p{#1}}
\newcolumntype{R}[1]{>{\PreserveBackslash\raggedleft}p{#1}}
\newcolumntype{L}[1]{>{\PreserveBackslash\raggedright}p{#1}}

\def\Dz {D^{0}}
\def\Db {\overline{D}{}^{0}}
\def\pip  {\pi^{+}}
\def\pim {\pi^{-}}

\def\dz {D^{0}}

\def\pip  {\pi^{+}}
\def\pim {\pi^{-}}
\def\ks {K_S^0}
\def\msp {m_+^2}
\def\msm {m_-^2}

\def\dzdecay {\Dz\to K_S^0\pip\pim}
\def\dsdecay {D^{*+}\to \Dz\pi_s^+}
\def\afb  {{\cal A}_f}
\def\abf  {|{\cal A}_{\bar{f}}|}
\def\afc  {{\cal A}_{\bar{f}}}

\def\simlt{\mathrel{\lower2.5pt\vbox{\lineskip=0pt\baselineskip=0pt
          \hbox{$<$}\hbox{$\sim$}}}}
\def\simgt{\mathrel{\lower2.5pt\vbox{\lineskip=0pt\baselineskip=0pt
           \hbox{$>$}\hbox{$\sim$}}}}

\usepackage[english]{babel}
\begin{document}

\renewcommand{\thefootnote}{\fnsymbol{footnote}}

\title{ \quad\\[1.0cm]  Measurement of ${D^{0}}$-${\overline{D}{}^{0}}$ Mixing and Search for Indirect $CP$ Violation Using ${D^0\to K_S^0\pi^+\pi^-}$  Decays}

\noaffiliation
\affiliation{University of the Basque Country UPV/EHU, 48080 Bilbao}
\affiliation{Beihang University, Beijing 100191}
\affiliation{University of Bonn, 53115 Bonn}
\affiliation{Budker Institute of Nuclear Physics SB RAS and Novosibirsk State University, Novosibirsk 630090}
\affiliation{Faculty of Mathematics and Physics, Charles University, 121 16 Prague}
\affiliation{Chiba University, Chiba 263-8522}
\affiliation{University of Cincinnati, Cincinnati, Ohio 45221}
\affiliation{Deutsches Elektronen--Synchrotron, 22607 Hamburg}
%%%\affiliation{Department of Physics, Fu Jen Catholic University, Taipei 24205}
\affiliation{Justus-Liebig-Universit\"at Gie\ss{}en, 35392 Gie\ss{}en}
%%%\affiliation{Gifu University, Gifu 501-1193}
%%%\affiliation{II. Physikalisches Institut, Georg-August-Universit\"at G\"ottingen, 37073 G\"ottingen}
%%%\affiliation{The Graduate University for Advanced Studies, Hayama 240-0193}
%%%\affiliation{Gyeongsang National University, Chinju 660-701}
\affiliation{Hanyang University, Seoul 133-791}
\affiliation{University of Hawaii, Honolulu, Hawaii 96822}
\affiliation{High Energy Accelerator Research Organization (KEK), Tsukuba 305-0801}
%%%\affiliation{Hiroshima Institute of Technology, Hiroshima 731-5193}
\affiliation{IKERBASQUE, Basque Foundation for Science, 48011 Bilbao}
%%%\affiliation{University of Illinois at Urbana-Champaign, Urbana, Illinois 61801}
\affiliation{Indian Institute of Technology Guwahati, Assam 781039}
\affiliation{Indian Institute of Technology Madras, Chennai 600036}
%%%\affiliation{Indiana University, Bloomington, Indiana 47408}
\affiliation{Institute of High Energy Physics, Chinese Academy of Sciences, Beijing 100049}
\affiliation{Institute of High Energy Physics, Vienna 1050}
\affiliation{Institute for High Energy Physics, Protvino 142281}
%%%\affiliation{Institute of Mathematical Sciences, Chennai 600113}
\affiliation{INFN - Sezione di Torino, 10125 Torino}
\affiliation{Institute for Theoretical and Experimental Physics, Moscow 117218}
\affiliation{J. Stefan Institute, 1000 Ljubljana}
\affiliation{Kanagawa University, Yokohama 221-8686}
\affiliation{Institut f\"ur Experimentelle Kernphysik, Karlsruher Institut f\"ur Technologie, 76131 Karlsruhe}
\affiliation{Kavli Institute for the Physics and Mathematics of the Universe (WPI), University of Tokyo, Kashiwa 277-8583}
%%%\affiliation{Department of Physics, Faculty of Sciences, King Abdulaziz University, Jeddah 21589}
\affiliation{Korea Institute of Science and Technology Information, Daejeon 305-806}
\affiliation{Korea University, Seoul 136-713}
%%%\affiliation{Kyoto University, Kyoto 606-8502}
\affiliation{Kyungpook National University, Daegu 702-701}
\affiliation{\'Ecole Polytechnique F\'ed\'erale de Lausanne (EPFL), Lausanne 1015}
\affiliation{Faculty of Mathematics and Physics, University of Ljubljana, 1000 Ljubljana}
\affiliation{Luther College, Decorah, Iowa 52101}
\affiliation{University of Maribor, 2000 Maribor}
\affiliation{Max-Planck-Institut f\"ur Physik, 80805 M\"unchen}
\affiliation{School of Physics, University of Melbourne, Victoria 3010}
\affiliation{Moscow Physical Engineering Institute, Moscow 115409}
\affiliation{Moscow Institute of Physics and Technology, Moscow Region 141700}
\affiliation{Graduate School of Science, Nagoya University, Nagoya 464-8602}
\affiliation{Kobayashi-Maskawa Institute, Nagoya University, Nagoya 464-8602}
%%%\affiliation{Nara University of Education, Nara 630-8528}
\affiliation{Nara Women's University, Nara 630-8506}
\affiliation{National Central University, Chung-li 32054}
\affiliation{National United University, Miao Li 36003}
\affiliation{Department of Physics, National Taiwan University, Taipei 10617}
\affiliation{H. Niewodniczanski Institute of Nuclear Physics, Krakow 31-342}
\affiliation{Nippon Dental University, Niigata 951-8580}
\affiliation{Niigata University, Niigata 950-2181}
\affiliation{University of Nova Gorica, 5000 Nova Gorica}
\affiliation{Osaka City University, Osaka 558-8585}
%%%\affiliation{Osaka University, Osaka 565-0871}
\affiliation{Pacific Northwest National Laboratory, Richland, Washington 99352}
\affiliation{Panjab University, Chandigarh 160014}
\affiliation{Peking University, Beijing 100871}
\affiliation{University of Pittsburgh, Pittsburgh, Pennsylvania 15260}
\affiliation{Punjab Agricultural University, Ludhiana 141004}
%%%\affiliation{Research Center for Electron Photon Science, Tohoku University, Sendai 980-8578}
%%%\affiliation{Research Center for Nuclear Physics, Osaka University, Osaka 567-0047}
\affiliation{RIKEN BNL Research Center, Upton, New York 11973}
%%%\affiliation{Saga University, Saga 840-8502}
\affiliation{University of Science and Technology of China, Hefei 230026}
%%%\affiliation{Seoul National University, Seoul 151-742}
%%%\affiliation{Shinshu University, Nagano 390-8621}
\affiliation{Soongsil University, Seoul 156-743}
\affiliation{Sungkyunkwan University, Suwon 440-746}
\affiliation{School of Physics, University of Sydney, NSW 2006}
\affiliation{Department of Physics, Faculty of Science, University of Tabuk, Tabuk 71451}
\affiliation{Tata Institute of Fundamental Research, Mumbai 400005}
\affiliation{Excellence Cluster Universe, Technische Universit\"at M\"unchen, 85748 Garching}
\affiliation{Toho University, Funabashi 274-8510}
\affiliation{Tohoku Gakuin University, Tagajo 985-8537}
\affiliation{Tohoku University, Sendai 980-8578}
\affiliation{Department of Physics, University of Tokyo, Tokyo 113-0033}
\affiliation{Tokyo Institute of Technology, Tokyo 152-8550}
\affiliation{Tokyo Metropolitan University, Tokyo 192-0397}
\affiliation{Tokyo University of Agriculture and Technology, Tokyo 184-8588}
\affiliation{University of Torino, 10124 Torino}
%%%\affiliation{Toyama National College of Maritime Technology, Toyama 933-0293}
\affiliation{CNP, Virginia Polytechnic Institute and State University, Blacksburg, Virginia 24061}
\affiliation{Wayne State University, Detroit, Michigan 48202}
\affiliation{Yamagata University, Yamagata 990-8560}
\affiliation{Yonsei University, Seoul 120-749}
\author{T.~Peng}\affiliation{University of Science and Technology of China, Hefei 230026} % USTC
\author{Z.~P.~Zhang}\affiliation{University of Science and Technology of China, Hefei 230026} % USTC
  \author{A.~Abdesselam}\affiliation{Department of Physics, Faculty of Science, University of Tabuk, Tabuk 71451} % Tabuk
  \author{I.~Adachi}\affiliation{High Energy Accelerator Research Organization (KEK), Tsukuba 305-0801} % KEK
% \author{K.~Adamczyk}\affiliation{H. Niewodniczanski Institute of Nuclear Physics, Krakow 31-342} % Krakow
  \author{H.~Aihara}\affiliation{Department of Physics, University of Tokyo, Tokyo 113-0033} % Tokyo
% \author{S.~Al~Said}\affiliation{Department of Physics, Faculty of Science, University of Tabuk, Tabuk 71451}\affiliation{Department of Physics, Faculty of Science, King Abdulaziz University, Jeddah 21589) % Tabuk
  \author{K.~Arinstein}\affiliation{Budker Institute of Nuclear Physics SB RAS and Novosibirsk State University, Novosibirsk 630090} % BINP
% \author{Y.~Arita}\affiliation{Graduate School of Science, Nagoya University, Nagoya 464-8602} % Nagoya
  \author{D.~M.~Asner}\affiliation{Pacific Northwest National Laboratory, Richland, Washington 99352} % PNNL
% \author{T.~Aso}\affiliation{Toyama National College of Maritime Technology, Toyama 933-0293} % Toyama
  \author{V.~Aulchenko}\affiliation{Budker Institute of Nuclear Physics SB RAS and Novosibirsk State University, Novosibirsk 630090} % BINP
  \author{T.~Aushev}\affiliation{Institute for Theoretical and Experimental Physics, Moscow 117218} % ITEP
  \author{R.~Ayad}\affiliation{Department of Physics, Faculty of Science, University of Tabuk, Tabuk 71451} % Tabuk
% \author{T.~Aziz}\affiliation{Tata Institute of Fundamental Research, Mumbai 400005} % Tata
% ...
% Ljubljana
% \author{N.~Zwahlen}\affiliation{\'Ecole Polytechnique F\'ed\'erale de Lausanne (EPFL), Lausanne 1015} % Lausanne
% \author{O.~Zyukova}\affiliation{Budker Institute of Nuclear Physics SB RAS and Novosibirsk State University, Novosibirsk 630090} % BINP
\collaboration{The Belle Collaboration}

\begin{abstract}
                 We report a  measurement of ${D^{0}}$-${\overline{D}{}^{0}}$ mixing parameters and a search for indirect $CP$ violation through a time-dependent
 amplitude analysis of ${D^0\to K_S^0\pi^+\pi^-}$ decays.
        The results are based on 921~fb$^{-1}$ of data accumulated with the Belle detector at the KEKB asymmetric-energy $e^+e^-$ collider.
     Assuming $CP$ conservation, we measure the mixing parameters $x=(0.56\pm0.19^{+0.03}_{-0.09}{^{+0.06}_{-0.09}})\%$ and $y=(0.30\pm0.15^{+0.04}_{-0.05}{^{+0.03}_{-0.06})}\%$,  where the errors are statistical, experimental systematic, and systematic due to the amplitude model, respectively. With $CP$ violation allowed, the parameters $|q/p|=0.90^{+0.16}_{-0.15}{^{+0.05}_{-0.04}}{^{+0.06}_{-0.05}}$ and $\arg(q/p)=(-6\pm11{\pm3}{^{+3}_{-4}})^{\circ}$ are found to be
consistent with conservation of $CP$ symmetry in mixing and in the interference between mixing and
decay, respectively.
\end{abstract}


\maketitle

\tighten

{\renewcommand{\thefootnote}{\fnsymbol{footnote}}}
\setcounter{footnote}{0}

The mixing rate and size of $CP$ Violation ($CPV$) in the charm sector are predicted to be very small in the standard model (SM) \cite{theory,theory2}. Thus, the measurements of $\Dz$-$\Db$ mixing and $CPV$ are sensitive probes of possible contributions
from  physics beyond the SM~\cite{Grossman:2006jg, Golowich:2007ka}.
Several studies show evidence or observation of the mixing phenomenon in the $\Dz$-$\Db$ system, while $CPV$ is not yet observed~\cite{Marko,babarkpi2007,cernob,cdf2007, babar2013, aubert2009}. In our study, the direct determination of $\Dz$-$\Db$ mixing  and $CPV$ are achieved simultaneously by exploring the time-dependent decay rate of self-conjugated $\dzdecay$ decays.

The particle-antiparticle mixing phenomenon causes an initially produced
(at proper time $t=0$) pure $D^0$ or $\overline{D}{}^0$ meson state to evolve
in time to a linear combination of $D^0$ and $\overline{D}{}^0$ states.
We describe the decay amplitudes for a $\Dz$ or a $\Db$ into the final state $K_S^0\pip\pim$, ${\cal A}_f$ ($\overline{\cal A}_f$), as a function of  the Dalitz plot (DP) variables $(\msp,\msm)=(m^2_{K_S^0\pip},m^2_{K_S^0\pim})$.
If $CP$ symmetry in the decays is assumed, i.e., ${\overline{\cal A}_f} = {\cal A}_{\overline{f}}={\cal A}(\msm,\msp)$,
we can derive the time dependent decay rates for $D^0$ and $\overline{D}{}^0$ decays
to the final state $f$ as~\cite{cleo2005}:
\begin{equation}\label{eq1}
  \begin{array}{ll}
  &{|{\cal M}(f, t)|}^2 =\frac{e^{-\Gamma t}}{2}
  \{  ({|{\cal A}_f|} ^2+|\frac{q}{p}|^{2} {|{\cal A}_{\bar{f}}|}^2)\cosh(\Gamma yt)
 \\ & \quad +
  ({|{\cal A}_f|} ^2-|\frac{q}{p}|^{2} {|{\cal A}_{\bar{f}}|}^2)\cos(\Gamma xt)\,\,\, \\&\quad+2\Re(\frac{q}{p}{\cal A}_{\bar{f}}{{\cal A}_f}^*)\sinh(\Gamma yt) -2\Im(\frac{q}{p}{\cal A}_{\bar{f}}{{\cal A}_f}^*)\sin(\Gamma xt) \},
  \end{array}
\end{equation}
\begin{equation}\label{eq2}
  \begin{array}{ll}
  &{|\overline{{\cal M}}(f, t)|}^2 =\frac{e^{-\Gamma t}}{2}
  \{  ({\abf} ^2+|\frac{p}{q}|^{2} {|\afb|}^{2})\cosh(\Gamma yt)
  \\&\quad+
  ({\abf} ^2-|\frac{p}{q}|^{2} {|\afb|}^{2})\cos(\Gamma xt)\,\,\,\\&\quad+2\Re(\frac{p}{q}{\afb}{\afc}^*)\sinh(\Gamma yt)-2\Im(\frac{p}{q}{\afb}{\afc}^*)\sin(\Gamma xt) \},
  \end{array}
\end{equation}
where the two dimensionless parameters that describe the $\Dz$-$\Db$
mixing, $x$ and $y$, are related to the mass and width difference of the
two mass eigenstates $|D_{1,2}\rangle = p|D^0\rangle\pm
q|\overline{D}{}^0\rangle$: $x = \frac{m_1-m_2}{\Gamma}$, $y =
\frac{\Gamma_1-\Gamma_2}{2\Gamma}$. Here $\Gamma$ is the mean decay width, $\Gamma = \frac{\Gamma_1+\Gamma_2}{2}$. The coefficients $p$ and $q$ are
complex coefficients, satisfying $|p|^2+|q|^2=1$. The time evolution of
neutral $D$ meson decays is exponential with the lifetime
$\tau_{\dz}=1/\Gamma$, modulated by the mixing parameters $x$ and $y$.
  The possible $CPV$ can cause $q/p\neq 1$, which will be considered later. So a time-dependent amplitude analysis of self-conjugated decays allows a direct measurement of charm mixing parameters ($x$, $y$)  and a simultaneous search for the $CPV$ in mixing, in the decay and in interference between mixing and decay. This method was developed by CLEO~\cite{cleo2005} and extended by Belle~\cite{PaperLm} and Babar~\cite{babar}. In this paper, we report a measurement of mixing parameters $x$ and $y$ and parameters probing $CP$ violation in charm mixing and interference between mixing and the decay. The results of this analysis supersede the previous Belle results given in Ref.~\cite{PaperLm}.

We analyze  a data sample of 921 fb$^{-1}$  recorded at or near the $\Upsilon(nS)$ ($n=4,5$) resonances produced at the KEKB collider \cite{kekb} and collected with the Belle detector \cite{belle}. The detector is a large-solid-angle magnetic spectrometer consisting of a silicon vertex detector (SVD), a 50-layer central drift chamber (CDC) for charged particle tracking and specific ionization measurement ($dE/dx$), an array of aerogel threshold Cherenkov counters (ACC), time-of-flight  scintillation counters (TOF),  and an array of CsI(Tl) crystals for electromagnetic calorimetry (ECL) located inside a superconducting solenoid coil that provides a 1.5 T magnetic field. An iron flux return located outside the coil is instrumented to detect $K^0_L$ mesons and identify muons (KLM). Two inner detector configurations were used. A 2.0 cm diameter beampipe and a 3-layer silicon vertex detector were used for the first sample of 156 fb$^{-1}$, while a 1.5 cm diameter beampipe, a 4-layer silicon detector and a small-cell inner drift chamber were used to record the remaining 765 fb$^{-1}$.

We reconstruct the $\Dz$ mesons through the decay chain $\dsdecay$ and $\dzdecay$ \cite{foo}, where $\pi_s$ is referred to as the slow pion. The charge of $\pi_s$ is used to tag the flavor of the $D$ meson.  We use information from ACC, TOF and CDC to perform likelihood-based particle identification in order to select  pions from $\Dz$ decays. The $\ks$ candidates are  reconstructed in the  $\pi^+ \pi^-$ final state; we require that the pion candidates are from a common vertex displaced from the $e^+e^-$ interaction point (IP) and have an invariant mass within $\pm$10~MeV/$c^2$ of the nominal $K^0_S$ mass \cite{pdg}.  The $\Dz$ candidates are reconstructed by combining each $\ks$ candidate with two oppositely charged pion candidates. These pions are required to have at least two SVD hits  each in the $z$ and azimuthal projections. A $D^{*+}$ candidate is reconstructed by combining the $\Dz$ candidate with a low-momentum charged track (the $\pi^+_s$ candidate). To suppress the combinatorial background and $B\overline{B}$ events, we require the $D^{*+}$ momentum in the center-of-mass (CM) frame to be greater than 2.5~GeV/$c$ and 3.1~GeV/$c$ for
the data taken at the CM energy of $\Upsilon(4S)$ and $\Upsilon(5S)$ mass, respectively.

The proper decay time of the $\Dz$ candidate is calculated by projecting
the vector joining the production and decay vertices, $\vec{l}$, onto the $\Dz$  momentum vector $\vec{p}_D$ in the lab frame: $t = (m_{D^0}/p_D)\vec{l} \cdot (\vec{p}_D/p_D)$, where $m_{\Dz}$ is the nominal $\Dz$  mass. The $D^0$ decay position is determined by fitting the two prompt charged tracks to a common vertex and
the $\Dz$ production point is taken to be the intersection of the trajectory of the $\Dz$ candidate with the IP region. We constrain the $\pi_s$ to originate from the obtained $\Dz$ production vertex. The sum of the fit-quality values for the vertex fits is required to be lower than 100. The uncertainty of the proper decay time ($\sigma_t$) is evaluated from the corresponding covariance matrices. We require $\sigma_t <$  1~ps to remove events with a poorly determined decay time (the maximum of the $\sigma_t$ distribution is at~0.15 ps).

\begin{figure}[h]
\includegraphics[width=0.25\textwidth]{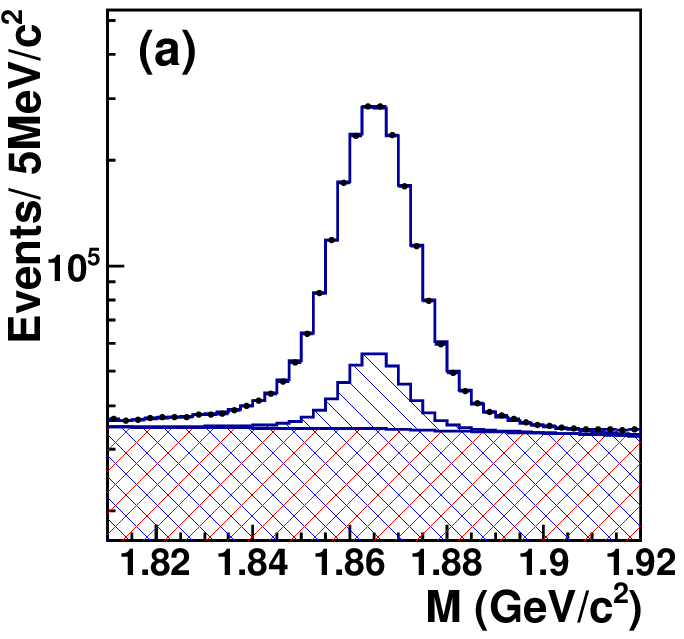}%
\includegraphics[width=0.25\textwidth]{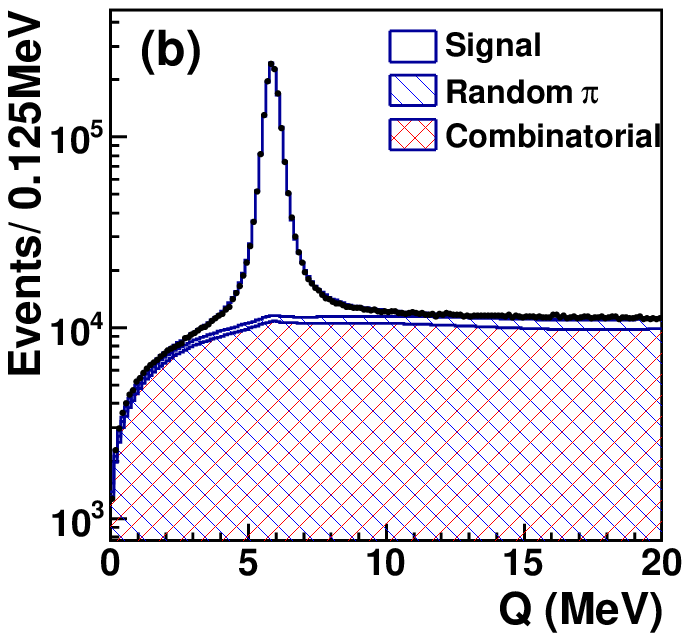}\\
\includegraphics[width=0.25\textwidth]{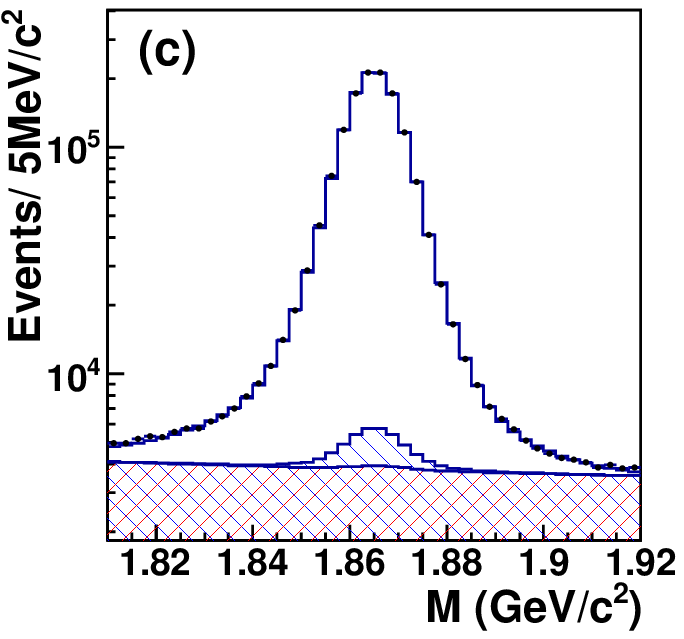}%
\includegraphics[width=0.25\textwidth]{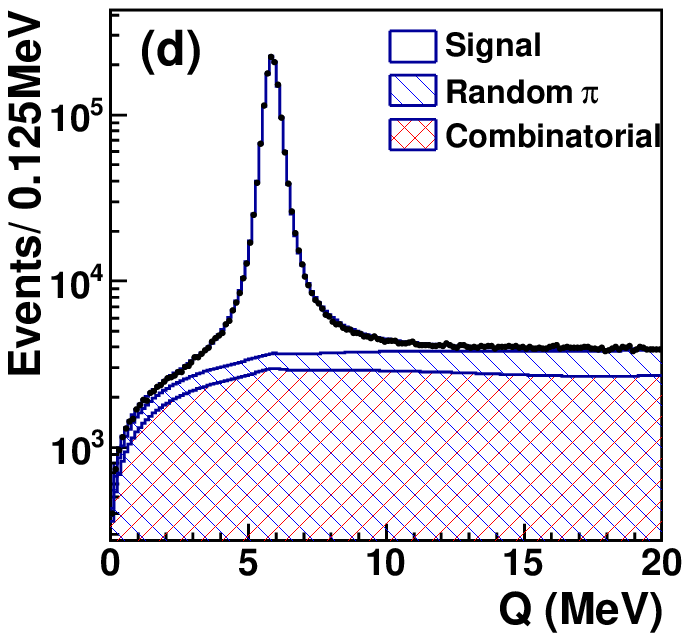}
\caption{\label{fitmq} The distributions of $M\equiv M_{K_S^0\pip\pim}$ and $Q \equiv (M_{K_S^0\pip\pim\pi_s}- M_{K_S^0\pip\pim}-m_{\pi_s})\cdot c^2$. (a) The projections of $M$
 for data (points with error bars) and $M-Q$ fit in the region 1.81~GeV/$c^2<M<1.92$~GeV/$c^2$ and $0<Q<20$~MeV including signal, random $\pi_s$
and combinatorial background. (b) The projections of  $Q$ in the region 1.81~GeV/$c^2<M<1.92$~GeV/$c^2$ and $0<Q<20$~MeV. (c) The corresponding projection of $M$ in the $Q$ signal region (5.75 MeV $<Q<5.95$ MeV). (d) The corresponding projection of $Q$ in the $M$ signal region ($|M-m_{D^0}|<15$~MeV/$c^2$). }
\end{figure}

We select events satisfying 1.81~GeV/$c^2$ $<M<$ 1.92~GeV/$c^2$ and 0 $< Q <$
20~MeV, where $M=M_{K_S^0\pip\pim}$ and $Q=(M_{K_S^0\pip\pim\pi_s}- M_{K_S^0\pip\pim}-m_{\pi_s})\cdot c^2$ are the
$\Dz$ invariant mass and kinetic energy released in the $D^*$ decay, respectively. About 3\% of selected events have two or more $D^*$ candidates. We select the best candidate as the one with the lowest fit-quality sum for the vertex fits. The $M$ and $Q$ distributions of the selected candidates are shown in Fig.~\ref{fitmq}.

We determine the signal yield from a two-dimensional fit to the $M-Q$ distribution. We parameterize the signal shape by a triple-Gaussian function for  $M$ and  the sum of a bifurcated Student's t-function and a Gaussian function for $Q$. We take the correlation between $M$ and $Q$ into account by parameterizing $\sigma_Q$ of the Student's t-function for $Q$ as a second-order polynomial in $|M-\mu_M|$ with $\mu_M$ being the mean of the Gaussian distribution for $M$. We include an additional term to describe  0.5\% of the signal candidates with a considerable amount of final state radiation. The backgrounds are classified into two types: random $\pi_s$ background, in which a random $\pi_s$ is combined with a true $\Dz$ candidate, and combinatorial background. The shape of the $M$ distribution for the random $\pi_s$ background is fixed to be the same as that used for the signal. Other background distributions are obtained from Monte Carlo (MC) simulation. The fit results are shown in Fig.~\ref{fitmq}. The small peaking components in the $Q$ distribution of combinatorial background are misreconstructed $\Dz$ decays with missing daughters. We find 1231731$\pm$1633(stat.) signal candidates with a purity of 95.5\% in the signal region defined as $|M-m_{D^0}|<15$~MeV/$c^2$, 5.75 MeV $<Q<5.95$ MeV.
\begin{table}[!hbtp]
\center \caption{\label{alldalitz}   Results for $\dzdecay$ Dalitz-plot parameters obtained from the mixing fit, including complex amplitudes, $\pi \pi$ S-wave and  $\ks\pi$ S-wave parameters, and fit fractions for each intermediate component. The errors are statistical only. }
\begin{tabular}{lcccc}\hline\hline
Resonance & Amplitude& Phase (deg) &Fit fraction\\
\hline
$K^*(892)^-$   &$1.590\pm    0.003$&$131.8\pm    0.2$  &0.6045\\
$K_0^*(1430)^-$&$2.059\pm    0.010$&$-194.6\pm    1.7$&0.0702\\
$K_2^*(1430)^-$&$1.150\pm    0.009$&$-41.5\pm    0.4$&0.0221\\
$K^*(1410)^-$  &$0.496\pm    0.011$&$83.4\pm    0.9$&0.0026\\
$K^*(1680)^-$  &$1.556\pm    0.097$&$-83.2\pm    1.2$&0.0016\\
$K^*(892)^+$   &$0.139\pm    0.002$&$-42.1\pm    0.7$&0.0046\\
$K_0^*(1430)^+$&$0.176\pm    0.007$&$ -102.3\pm    2.1$&0.0005\\
$K_2^*(1430)^+$&$0.077\pm    0.007$&$-32.2\pm    4.7$&0.0001\\
$K^*(1410)^+$  &$0.248\pm    0.010$&$-145.7\pm    2.9$&0.0007\\
$K^*(1680)^+$  &$1.407\pm    0.053$&$86.1\pm    2.7$&0.0013\\
\hline
$\rho(770)$& 1 (fixed)& 0 (fixed)&0.2000\\
$\omega(782)$&$0.0370\pm    0.0004$&$114.9\pm    0.6$&0.0057\\
$f_2(1270)$  &$1.300\pm    0.013$&$-31.6\pm    0.5$ & 0.0141\\
$\rho(1450)$ &$0.532\pm    0.027$&$80.8\pm    2.1$  & 0.0012\\
\hline
$\pi\pi$ S-wave&&&0.1288\\
$\beta_1$&$4.23\pm    0.02$&$164.0\pm    0.2$\\
$\beta_2$& $10.90\pm    0.02$&$15.6\pm    0.2 $ \\
$\beta_3$& $37.4\pm    0.3$&$3.3\pm   0.4$ \\
$\beta_4$& $14.7\pm    0.1$&$-8.9\pm    0.3$\\
$f_{11}^{\rm prod}$ &$12.76\pm    0.05$&$-161.1\pm    0.3$\\
$f_{12}^{\rm prod}$ &$14.2\pm    0.2$&$ -176.2\pm    0.6$\\
$f_{13}^{\rm prod}$ &$10.0\pm    0.5$&$-124.7\pm    2.1$\\
\hline
$K\pi$ S-wave & Parameters\\
M(MeV/$c^2$)  &$1461.7\pm    0.8$\\
$\Gamma$(MeV/$c^2$)&$268.3\pm    1.1$\\
F      &$0.4524\pm    0.005$          \\
$\phi_F(rad)$   &$0.248\pm    0.003$ \\
R              &1(fixed)  \\
$\phi_R(rad)$    &$2.495\pm    0.009$ \\
a(GeV/$c^{-1}$)  &$0.172\pm    0.006$\\
r(GeV/$c^{-1}$)  &$-20.6\pm    0.3$\\
\hline
$K^*(892)$ & Parameters\\
$M_{K^*{(892)}}$(MeV/$c^2$)  & $893.68\pm0.04$ \\
$\Gamma_{K^*{(892)}}$(MeV/$c^2$) & $47.49\pm0.06$ \\
\hline\hline
\end{tabular}
\end{table}

Mixing parameters are extracted from an unbinned maximum likelihood fit to $m_+^2$, $m_-^2$ and the decay time $t$ for the events selected in the signal region.
  The $\dzdecay$ decay rates  are expressed  in Eqs.~(\ref{eq1}) and (\ref{eq2}).  The reconstruction efficiency  over the DP plane is described by a cubic polynomial of $m_+^2$ and $m_-^2$ determined from a large MC sample of signal events.  The proper decay time resolution function is represented by a sum of three (two) Gaussians in the case of the 4-layer (3-layer) silicon vertex detector configuration.  We allow one of the Gaussians' means to differ from the other two for the case of the 4-layer  silicon vertex detector configuration.

 The Dalitz amplitudes ${\cal A}_f$ and $\bar{{\cal A}_f}$ are expressed as a sum of quasi-two-body amplitudes. For the P- and D-wave decays, we include 12 intermediate resonances described by relativistic Breit-Wigner parameterizations with mass dependent widths, Blatt-Weisskopf penetration factors as form factors and Zemach tensors for the angular dependence \cite{dalitzfo}. For the $\pi \pi$ S-wave dynamics, we adopt the K-matrix formalism with $P$-vector approximation~\cite{babarkmatrix}. For the $\ks\pi$ S-wave, we follow the same description as in Ref.~\cite{babar}. We tested different decay amplitude models by adding or removing resonances with small contributions or by using alternative parameterizations.

The random $\pi_s$ background contains real $\Dz$  and $\Db$  candidates; for these events, the charge of the $\pi_s$  is uncorrelated
with the flavor of the neutral $D$. Thus the PDF is taken to be $(1-f_w)|{\cal M}(f, t)|^2 + f_w|\overline{{\cal M}}(f, t)|^2$, convolved with the same resolution
function as that used for the signal, where $f_w$ is the wrong-tagged fraction. We measure $f_w$ by performing a fit to the candidates that populate the $Q$ sideband 3~MeV $< |Q-5.85~$MeV$ | < 14.15$~MeV, resulting in $f_w=0.511 \pm 0.003$.
The DP and decay time PDFs for combinatorial background are determined from the $M$ sideband (30~MeV/$c^2$$<|M-m_{\dz}|<50$~MeV/$c^2$). The decay time PDF is described using the sum of a delta function and an exponential component convolved by a triple-Gaussian  as a resolution function.
We validate the fitting procedure with fully simulated MC experiments. The fitter returns the mixing parameters consistent with the inputs for signal samples with and without background events included.
\begin{figure}[h!]
\center
\includegraphics[width=0.25\textwidth]{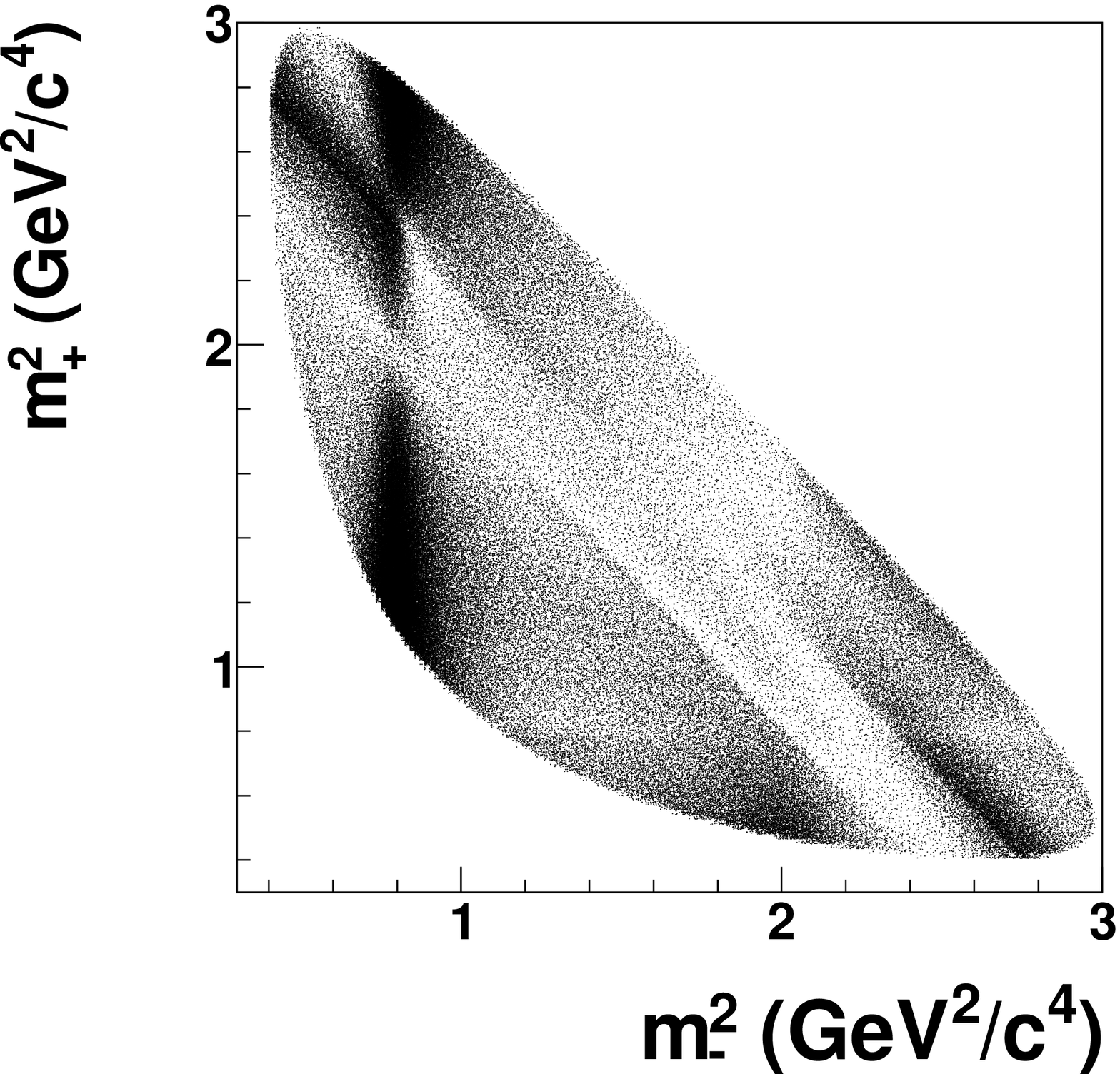}%
\includegraphics[width=0.25\textwidth]{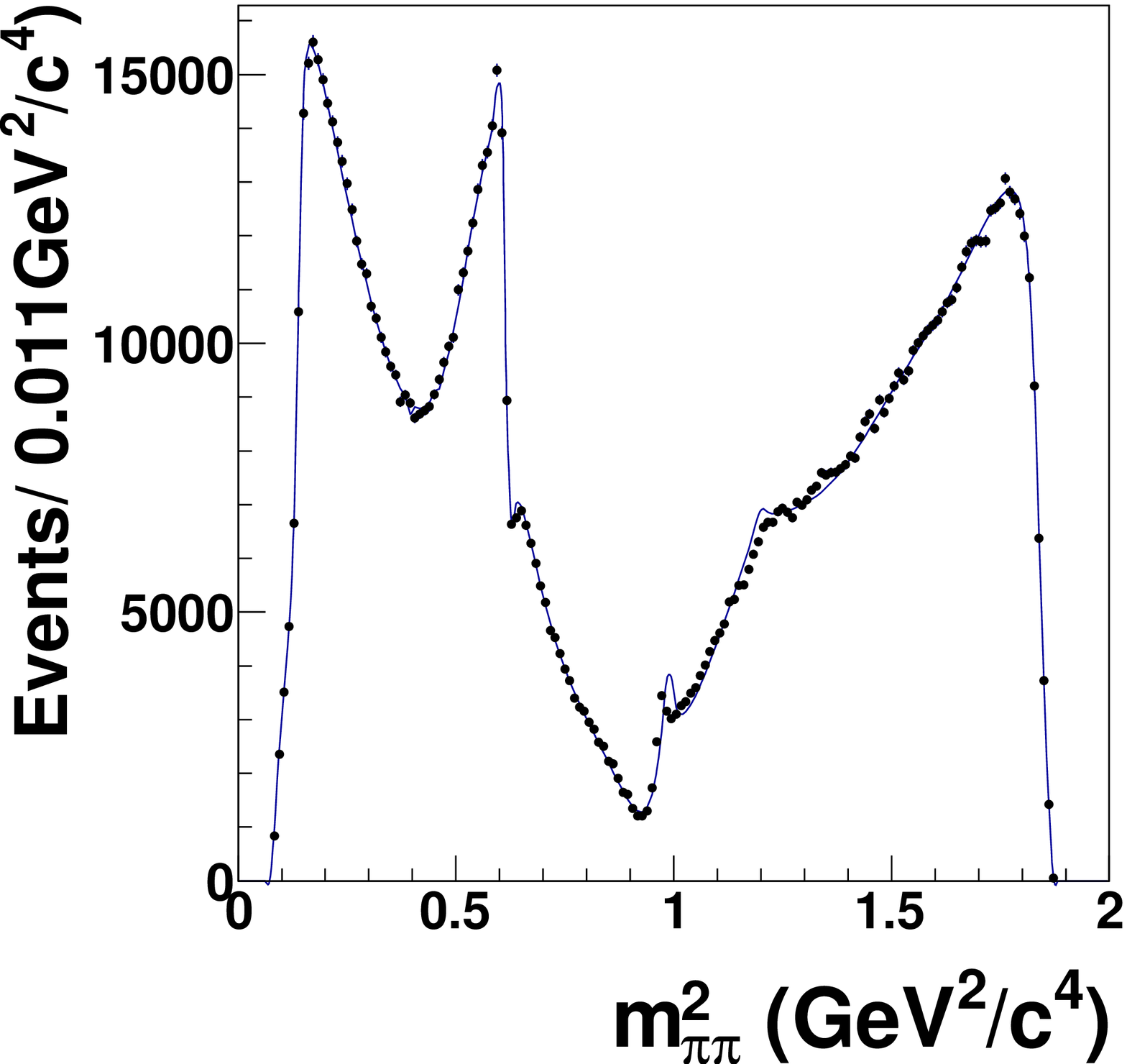}

\hspace*{300pt}

\includegraphics[width=0.25\textwidth]{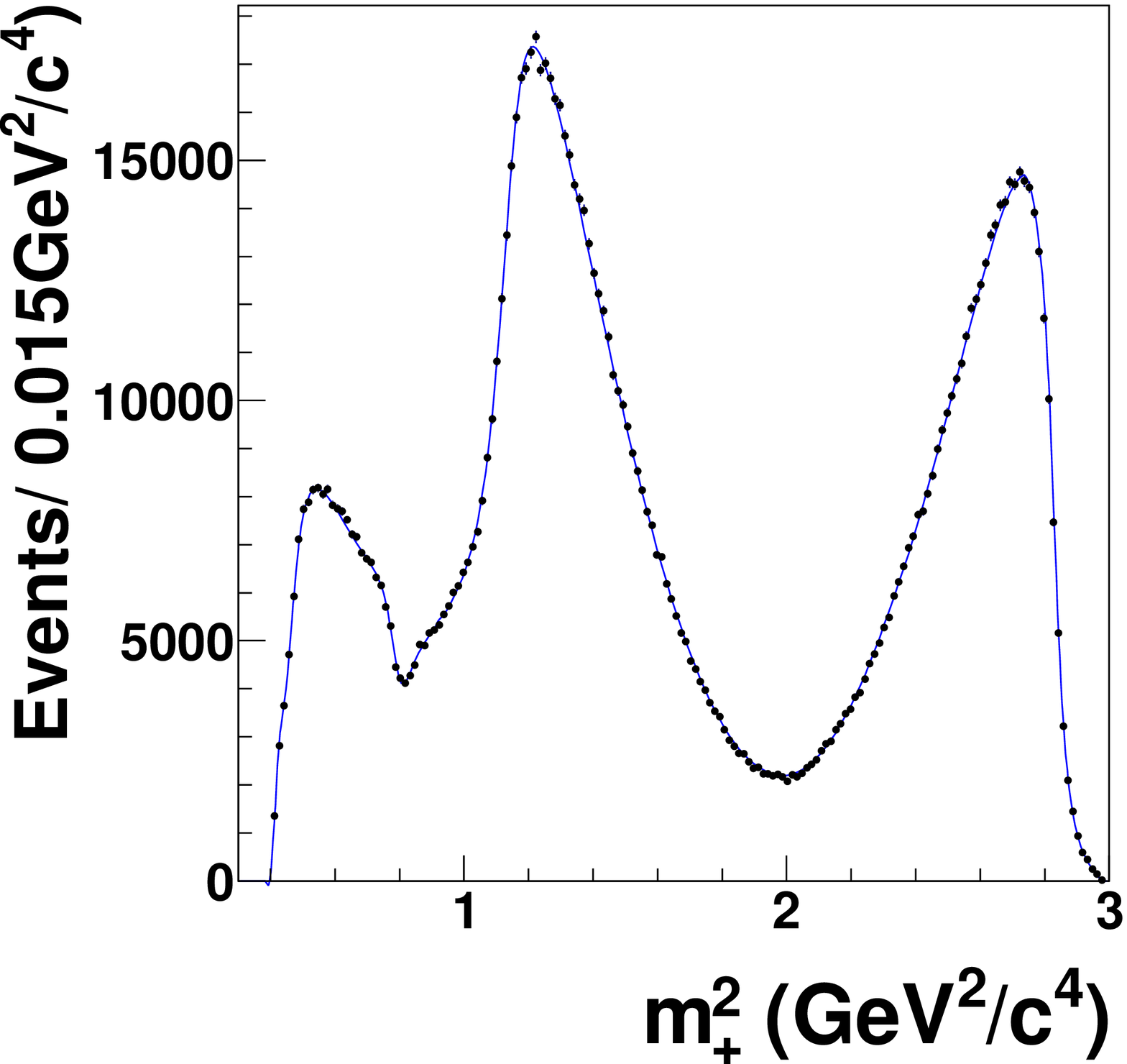}%
\includegraphics[width=0.25\textwidth]{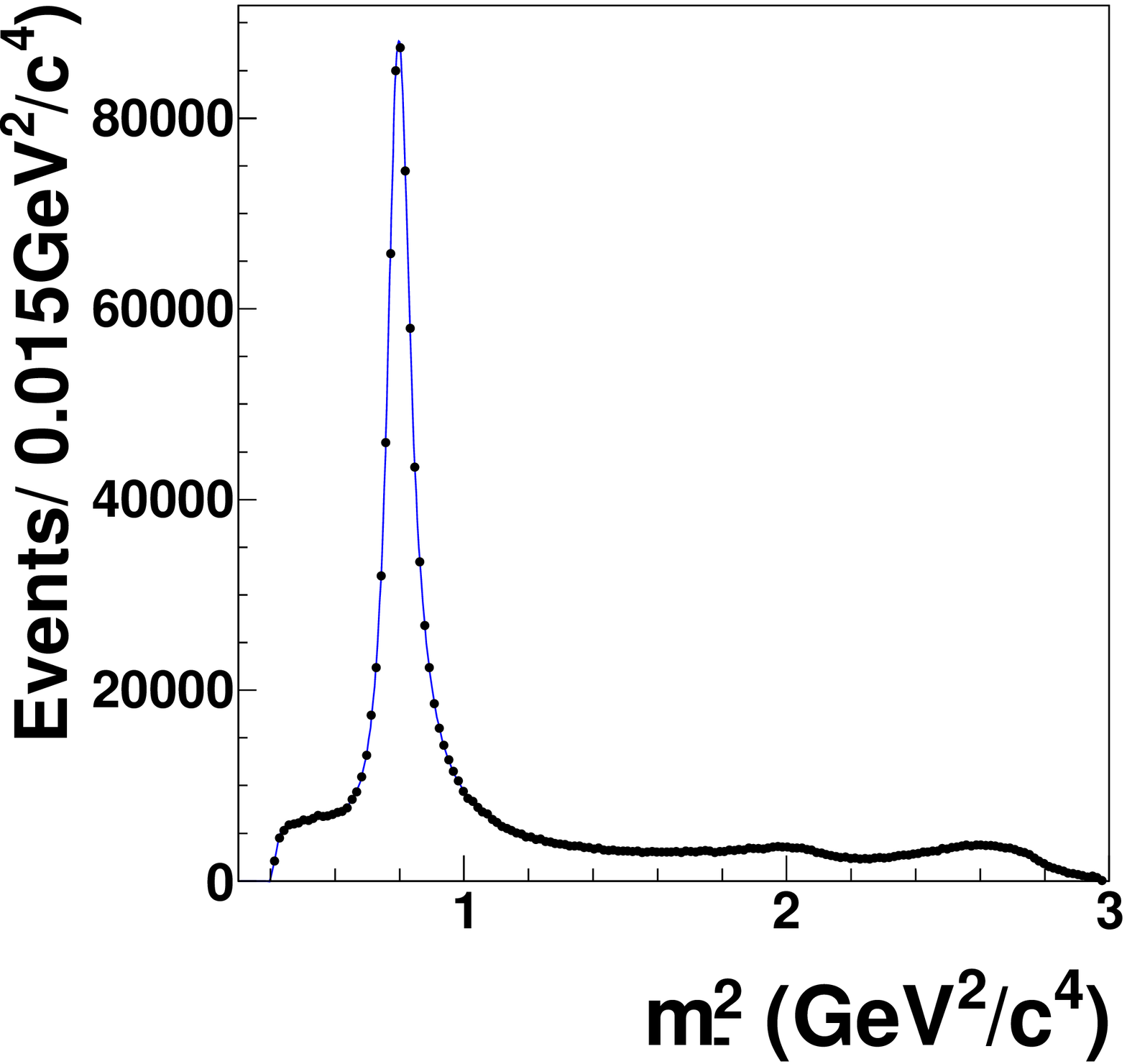}

%\break where the errors are statistical, experimental and model uncertainties respectively.
\caption{\label{dalitz}Dalitz distribution and Dalitz variables ($m_+^2$, $m_-^2$ and $m_{\pi\pi}^2$)
projections for the selected data sample. The full line represents the result of the fit
described in the text. }
\end{figure}

\begin{figure}[!htb]
\includegraphics[width=0.45\textwidth]{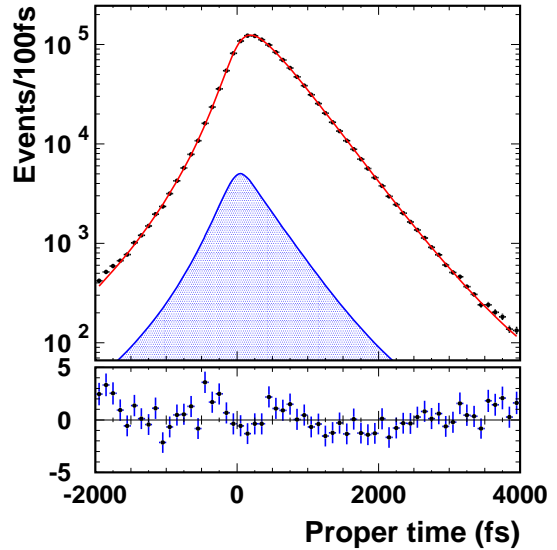}\break
\caption{\label{timeallt}  The proper-time distribution for events in the signal region (points) and
fit projection for the $CP$ conserved fit (curve). The shaded region shows the combinatorial components. The residuals are shown below the plot.}
\end{figure}

We first perform a decay-time integrated fit to the DP distribution by setting the amplitudes and phases for intermediate states free separately for $\Dz$ and $\Db$ decays. We observe  that the two sets of parameters are consistent and so, hereinafter, assume ${\overline{\cal A}_f} = {\cal A}_{\overline{f}}$.
In our subsequent fit to the data sample, we set the free parameters to be ($x$, $y$), the $\Dz$ lifetime $\tau$, the parameters of the proper decay time resolution function, and the amplitude model parameters. We extract the mixing parameters $x=(0.56\pm0.19)\%$ and $y=( 0.30\pm0.15)\%$, with the statistical correlation coefficient between $x$ and $y$ of 0.012. We also determine the $\Dz$ mean lifetime $\tau$ = (410.3$\pm$ 0.6) fs, in agreement with the world average~\cite{pdg}.
The projections of the DP distribution and $\Dz$ proper time are shown in Figs.~\ref{dalitz} and~\ref{timeallt}, respectively. Table \ref{alldalitz} lists the results for the DP resonance parameters. To evaluate the fit quality of the amplitude fit, we perform a two-dimensional  $\chi^2$ test over the DP plane, obtaining $\chi^2/ndf$=1.207 for $14264-49$ degrees of freedom ($ndf$).  The fit correctly reproduces the DP of the data, with some small discrepancies at the dips of the distribution in the central $m_{\pi\pi}^2$ region (1.0GeV$^2/c^4$ $<m_{\pi\pi}^2<$1.3GeV$^2/c^4$).
\renewcommand{\arraystretch}{1.2}
\begin{table}[htb]
\caption{\label{result}Fit results for the mixing parameters $x$ and $y$ from the $CP$-conserved fit and the $CPV$-allowed fit. The errors are statistical, experimental systematic, and systematic due to the amplitude model, respectively.}
\begin{tabular}{lccc}\hline
Fit type& Parameter &Fit result\\\hline
No $CPV$&$x(\%)$&$0.56\pm0.19^{+0.03}_{-0.09}{^{+0.06}_{-0.09}}$\\
&$y(\%)$&$0.30\pm0.15^{+0.04}_{-0.05}{^{+0.03}_{-0.06}}$\\
$CPV$&$x(\%)$&$0.56\pm0.19^{+0.04}_{-0.08}{^{+0.06}_{-0.08}}$\\
&$y(\%)$&$0.30\pm0.15^{+0.04}_{-0.05}{^{+0.03}_{-0.07}}$\\
&$|q/p|$&$0.90^{+0.16}_{-0.15}{^{+0.05}_{-0.04}}{^{+0.06}_{-0.05}}$\\
&$\arg(q/p)(^{\circ})$&$-6\pm11{\pm3}{^{+3}_{-4}}$\\\hline
\end{tabular}
\end{table}

We also search for $CPV$  in $\Dz/\Db\to K_S^0\pip\pim$ decays. The $CPV$ parameters $|q/p|$ and $\arg(q/p)$ are included in the PDF. The values for the mixing parameters from this fit are essentially identical to the ones from the $CP$-conserved fit. The resulting $CPV$ parameters are $|q/p|=0.90^{+0.16}_{-0.15}$ and $\arg(q/p)=(-6\pm11)^{\circ}$~\footnotemark \footnotetext{
The correlations among  the mixing and $CPV $parameters are: \center
\begin{tabular}{c|l*{4}{C{24pt}}}
                        \hline\hline
                        &\multicolumn{4}{c}{Correlation coefficient}\\
                          & $x$ & $y$ & $|q/p|$ &$\arg(q/p)$ \\\hline
                         $x$ & 1 & 0.054 & -0.074 &-0.031\\
                         $y$ &   & 1 & 0.034 &-0.019\\
                         $|q/p|$ &   &   & 1 &0.044\\
                         $\arg(q/p)$ &  &  &  &1\\
                         \hline\hline
                       \end{tabular}
}. The results from the two fits are listed in Table~\ref{result}.

We consider several contributions to the experimental systematic uncertainty, which are summarized in Table~\ref{sysexp}.
The uncertainty associated with best candidate selection is estimated by fitting a data sample that excludes all events with multiple candidates.  The uncertainties due to  signal and background yields determination are evaluated by varying their values by the corresponding statistical uncertainties. The uncertainties due to determination of the fraction of wrong tagged events in random $\pi_s$ background are estimated by letting the fraction parameter free  in the mixing fit, which leads to $f_w=0.44 \pm 0.02$. To account for the uncertainty associated with signal time resolution parameterization, we remove the offset in the third Gaussian function for the case of the 4-layer  silicon vertex detector configuration. The uncertainty associated with the DP efficiency function is estimated by replacing it with the second-order polynomial parameterization. The uncertainties due to the small misalignment of detectors are estimated to be negligible by varying the offset of the resolutions function.
The uncertainties associated with the combinatorial-background PDF are estimated by choosing different sideband samples to fit distributions and varying the PDF shape parameters by their statistical errors. To evaluate uncertainties associated with a possible correlation between the DP and the time distribution for the combinatorial background, we parameterize the DP distribution in different decay time intervals.  We also vary the ratios of certain DCS intermediate states and corresponding CF ones by estimated biases using simulated samples ($\sim5\%$) in order to estimate  uncertainties raised by the fitting procedure. The dominant contributions of experimental systematic error are from the determination of background PDFs and the DP's fitting procedure.
\begin{table*}[htb]
\center \caption{\label{sysexp}Summary of the contributions to experimental systematic uncertainty on the mixing and $CPV$ parameters. The positive and negative
errors are added in quadrature separately.}
\begin{tabular}{l|cc|cccc}\hline\hline
&\multicolumn{2}{c|}{No $CPV$}& \multicolumn{4}{c}{$CPV$}\\
Source & $\Delta x/ 10^{-4}$&$\Delta y/ 10^{-4}$ & $\Delta x/ 10^{-4}$&$\Delta y/ 10^{-4}$ &$|q/p|/ 10^{-2} $ & $\arg(q/p)/ ^{\circ}$\\\hline
Best candidate selection&$+1.0$&$+1.9$&$+1.3$&$+2.0$&$-2.3$&$+2.2$\\
Signal and background yields & $\pm$0.3&$\pm$0.3&$\pm$0.4&$\pm$0.4&$\pm$1.2&$\pm$0.8\\
Fraction of wrong tagged events& $-0.7$&$-0.4$& $-0.5$&$+0.4$& $+1.1$&$+0.8$\\
Time resolution of signal &  $-1.4$&$-0.9$& $-1.2$&$-0.8$& $+0.8$&$-1.2$\\
Efficiency& $-1.1$&$-2.1$& $-1.4$&$-2.2$& $+3.1$&$+1.3$\\
Combinatorial PDF &  $^{+1.9}_{-4.8}$&$^{+2.3}_{-3.9}$&$^{+2.4}_{-4.1}$&$^{+2.0}_{-4.4}$&$^{+1.2}_{-2.9}$&$^{+2.8}_{-2.3}$\\
$K^*(892)$ DCS/CF reduced by $5\%$&  $-7.3$& $+2.3$& $-6.9$&$+3.1$& $+3.3$&$-1.4$\\
$K_2^*(1430)$ DCS/CF reduced by $5\%$& +1.7& $-0.7$& $+2.2$&$-0.2$& $+1.1$&$+0.4$\\\hline
Total&  $^{+2.8}_{-8.9}$  &$^{+3.7}_{-4.6}$ &$^{+3.6}_{-8.3}$&$^{+4.3}_{-5.1}$  &$^{+5.0}_{-4.0}$&$^{+3.3}_{-3.0}$ \\\hline\hline
\end{tabular}
\end{table*}

\begin{table*}[htb]
\center \caption{\label{sysmod}Summary of contributions to the modeling systematic uncertainty on the mixing and $CPV$ parameters. The positive and negative
errors are added in quadrature separately.}
\begin{tabular}{l|cc|cccc}\hline\hline
&\multicolumn{2}{c|}{No $CPV$}& \multicolumn{4}{c}{$CPV$}\\
Source& $\Delta x/ 10^{-4}$&$\Delta y/ 10^{-4}$ & $\Delta x/ 10^{-4}$&$\Delta y/ 10^{-4}$ &$|q/p|/ 10^{-2} $ & $\arg(q/p)/ ^{\circ} $\\\hline
Resonance M $\&$ $\Gamma$ &$\pm1.4$&$\pm1.2$&$\pm1.2$&$\pm1.3$&$\pm2.1$&$\pm1.0$\\
 $K^*(1680)^+$ removal&$-1.8$ &$-3.0$&$-2.2$ &$-2.8$&$+2.1$ &$-1.2$\\
 $K^*(1410)^{\pm}$ removal&$-1.2$&$-3.6$&$-1.7$ &$-3.9$&$-1.3$ &$+1.4$\\
 $\rho(1450)$ removal&+2.1&+0.3&$+2.1$ &$+0.5$&$-1.9$ &$+0.9$\\
Form factors  &  $+4.0$&$+2.4$&  $+4.3$&$+2.0$&  $-2.4$&$-1.0$\\
$\Gamma(q^2)={\rm constant}$& $+3.3$&$-1.6$&  $+4.1$&$-2.3$&  $-1.6$&$+1.3$\\
Angular dependence & $-8.5$&$-3.9$&  $-7.4$&$-3.6$&  $+5.6$&$-3.2$\\
K-matrix formalism& $-2.2$&$+1.8$&  $-3.5$&$+2.4$&  $-3.6$&$+1.1$\\\hline
Total&  $^{+5.8}_{-9.1}$  &$^{+3.2}_{-6.4}$  &$^{+6.4}_{-8.4}$&$^{+3.4}_{-6.9}$ &$^{+6.4}_{-5.1}$&$^{+2.5}_{-3.7}$\\\hline\hline
\end{tabular}
\end{table*}
We estimate uncertainties due to the Dalitz model assumptions by  repeating the fit with slightly different models as described below, and the results are summarized in Table~\ref{sysmod}.
We vary the average values of masses and widths for the included resonances by their uncertainties from~\cite{pdg}. We  remove the $K^*(1680)^+$, $K^*(1410)^{\pm}$ and $\rho(1450)$ resonances which contribute small fractions in the $\dzdecay$ channel.  We perform fits with no form factors and with constant Breit-Wigner widths. The uncertainty due to the angular distribution for intermediate states is estimated by trying an alternative helicity angular formalism \cite{dalitzfo}. We replace the model for  $\pi \pi$ S-wave of DP by a different K-matrix formalism \cite{AS}. The main contributions are from the parameterizations of  angular dependence.
\begin{figure}[!htbp]
\includegraphics[width=0.4\textwidth]{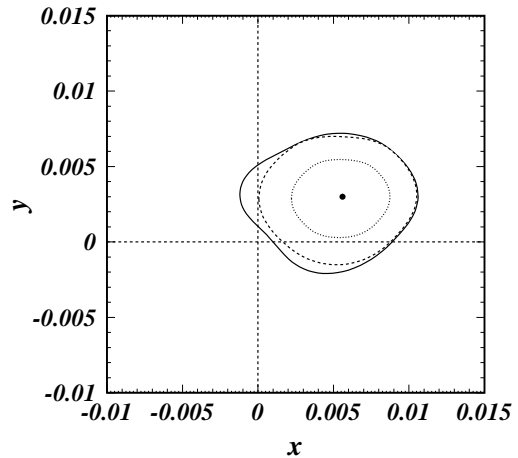}\break
\caption{\label{contsta} Central value (point) and   C.L. contours for ($x$, $y$): dotted (dashed) corresponds to  $68.3\%$ ($95\%$) C.L. contour for $CP$-conserved Dalitz fit, and solid corresponds to $95\%$ C.L. contour for $CPV$-allowed fit  with statistical, experimental and model uncertainties included.  }
\end{figure}
By exploring  the negative log-likelihood distribution on the plane of mixing parameters, we draw the two-dimensional ($x$, $y$) confidence-level (C.L.) contours  for both the $CP$-conserved and $CPV$-allowed fits (Fig.~\ref{contsta}). We combine the statistical and systematic uncertainties using the method described in~\cite{PaperLm}.

 In summary, an updated measurement of $\Dz$-$\Db$ mixing in $\dzdecay$ decays was performed based on 921 fb$^{-1}$ of
data collected with the Belle detector. The results supersede our results in Ref.~\cite{PaperLm}.
 We obtain $x=(0.56\pm0.19^{+0.03}_{-0.09} {^{+0.06}_{-0.09}})\%$, $y= (0.30\pm0.15^{+0.04}_{-0.05}{^{+0.03}_{-0.06}})\%$ assuming no $CPV$, where the errors are statistical, experimental systematic, and systematic due to the amplitude model, respectively.
  The significance of $\Dz$-$\Db$ mixing is estimated to be 2.5 standard deviations relative to the no-mixing point ($x=0$,~$y=0$). Comparing with previous measurements \cite{PaperLm,babar}, these results give a consistent determination of $\Dz$-$\Db$ mixing with significantly improved sensitivity.
A search for  $CP$ violation  results in the most accurate values of the $|q/p|$ and $\arg(q/p)$ parameters in a single experiment: $|q/p|=0.90^{+0.16}_{-0.15}{^{+0.05}_{-0.04}}{^{+0.06}_{-0.05}}$, $\arg(q/p)=(-6\pm11{\pm3}{^{+3}_{-4}})^{\circ}$. The values are consistent with no $CPV$.

We thank the KEKB group for excellent operation of the
accelerator; the KEK cryogenics group for efficient solenoid
operations; and the KEK computer group, the NII, and
PNNL/EMSL for valuable computing and SINET4 network support. We acknowledge support from MEXT, JSPS and Nagoya's TLPRC (Japan);
ARC and DIISR (Australia); FWF (Austria); NSFC (China); MSMT (Czechia);
CZF, DFG, and VS (Germany);
DST (India); INFN (Italy); MEST, NRF, GSDC of KISTI, and WCU (Korea);
MNiSW and NCN (Poland); MES and RFAAE (Russia); ARRS (Slovenia);
IKERBASQUE and UPV/EHU (Spain);
SNSF (Switzerland); NSC and MOE (Taiwan); and DOE and NSF (USA).


\begin{thebibliography}{99}
\bibitem{theory}A.A. Petrov, Int. J. Mod. Phys. A {\bf 21}, 5686 (2006).
\bibitem{theory2}A.F.~Falk,Y.~Grossman, Z.~Ligeti, A.~A.~Petrov, Phys.\ Rev.\ D {\bf 65}, 054034 (2002).
%\cite{Grossman:2006jg}
\bibitem{Grossman:2006jg}  Y.~Grossman, A.~L.~Kagan and Y.~Nir,
  Phys.\ Rev.\ D {\bf 75}, 036008 (2007).
\bibitem{Golowich:2007ka}
  E.~Golowich, J.~A.~Hewett, S.~Pakvasa and A.~A.~Petrov,
  Phys.\ Rev.\ D {\bf 76}, 095009 (2007).
    \bibitem{Marko}M.~Staric {\em et al.} (Belle Collaboration), Phys.\ Rev.\ Lett.\  {\bf 98}, 211803 (2007).
\bibitem{babarkpi2007}B.~Aubert {\it et al.} (BaBar Collaboration),
  Phys.\ Rev.\ Lett.\  {\bf 98}, 211802 (2007).
 \bibitem{aubert2009}B.~Aubert {\it et al.} (BaBar Collaboration),  Phys.\ Rev.\ Lett.\  {\bf 103}, 211801 (2009).
\bibitem{babar2013}J.~P.~Lees {\it et al.} (BaBar Collaboration),
  Phys.\ Rev.\ D {\bf 87}, 012004 (2013).
 \bibitem{cernob}R.~Aaij {\em et al.} (LHCb Collaboration), arXiv:1309.6534 [hep-ex].
 \bibitem{cdf2007}T.~Aaltonen {\it et al.} (CDF Collaboration), arXiv:1309.4078 [hep-ex].
 \bibitem{cleo2005}D.~M.~Asner {\em et al.} (CLEO Collaboration), Phys.\ Rev.\ D {\bf 72}, 012001 (2005); arXiv:hep-ex/0503045.
\bibitem{PaperLm}L.~M.~Zhang {\em et al.} (Belle Collaboration), Phys.\ Rev.\ Lett.\ {\bf 99}, 131803 (2007).
\bibitem{babar}B.~Aubert {\em et al.} (BaBar Collaboration), Phys.\ Rev.\ Lett.\ {\bf 105}, 081803 (2010).
\bibitem{kekb} S.~Kurokawa and E.~Kikutani, Nucl. Instrum. Methods Phys. Res. Sect.
 A {\bf 499}, 1 (2003), and other papers included in this Volume;
 T.Abe {\it et al.}, Prog. Theor. Exp. Phys. {\bf 2013}, 03A001 (2013) and following
 articles up to 03A011.
\bibitem{belle}A.~Abashian {\em et al.} (Belle Collaboration),  Nucl. Instrum. Methods Phys. Res., Sect. A {\bf 479}, 117 (2002); J.~Brodzicka {\em et al.}, Prog. Theor. Exp. Phys. 04D001 (2012).
\bibitem{foo} Charge conjugation is assumed throughout this paper unless stated otherwise.
\bibitem{pdg}J.~Beringer {\em et al.} (Particle Data Group), Phys.\ Rev.\ D {\bf 86}, 010001 (2012).
\bibitem{dalitzfo}For a review, see resonances' parameterizations in \cite{pdg}; M.~Jacob and G.~C.~Wick, Annals Phys.\ {\bf 7}, 404 (1959) [Annals Phys.\ {\bf 281}, 774 (2000)].
\bibitem{babarkmatrix}B.~Aubert {\em et al.} (BaBar Collaboration), Phys.\ Rev.\ D {\bf 78}, 034023 (2008).
\bibitem{AS} V.~V.~Anisovich and A.~V.~Sarantsev, Eur.~Phys.~J. A {\bf 16}, 229 (2003).

\end{thebibliography}
\end{document}